\documentclass{ws-ijmpd}

\input{epsf}

\usepackage{amssymb}
\usepackage{graphicx}

\def\eq#1{Eq. (\ref{#1})}

\def\al{\alpha} \def\be{\beta} \def\ga{\gamma} \def\de{\delta}
\def\ep{\epsilon}   
\def\th{\theta}   
\def\la{\lambda}

\def\om{\omega} \def\Ga{\Gamma} \def\De{\Delta} 
\def\La{\Lambda}   
 \def\Om{\Omega}

 \def\frac#1#2{{\textstyle{{#1}\over
{#2}}}} 
\def\lsim{\mathrel{\rlap{\lower4pt\hbox{\hskip1pt$\sim$}}
\raise1pt\hbox{$<$}}}
\def\gsim{\mathrel{\rlap{\lower4pt\hbox{\hskip1pt$\sim$}}
\raise1pt\hbox{$>$}}} \def\sqr#1#2{{\vcenter{\vbox{\hrule height.#2pt
\hbox{\vrule width.#2pt height#1pt \kern#1pt \vrule width.#2pt} \hrule
height.#2pt}}}}

\def\beq{\begin{equation}} \def\eeq{\end{equation}}
\def\beqa{\begin{eqnarray}} \def\eeqa{\end{eqnarray}}

\long\def\symbolfootnote[#1]#2{\begingroup%
\def\thefootnote{\fnsymbol{footnote}}\footnote[#1]{#2}\endgroup}

\begin{document}

\markboth{O. Bertolami and J. P\'aramos}
{Using global positioning systems to test extensions of General Relativity}

%%%%%%%%%%%%%%%%%%%%% Publisher's Area please ignore %%%%%%%%%%%%%%%
%
\catchline{}{}{}{}{}
%
%%%%%%%%%%%%%%%%%%%%%%%%%%%%%%%%%%%%%%%%%%%%%%%%%%%%%%%%%%%%%%%%%%%%

\title{Using global positioning systems to test extensions of General Relativity}

\author{ORFEU BERTOLAMI\footnote{Also at Instituto de Plasmas e F\'isica Nuclear, Instituto Superior T\'ecnico, Avenida Rovisco Pais 1, 1049-001 Lisboa, Portugal.}}

\address{Departamento de F\'{\i}sica e Astronomia, Faculdade de Ci\^encias, Universidade do Porto,\\Rua do Campo Alegre 687,
4169-007 Porto, Portugal\\
orfeu.bertolami@fc.up.pt}

\author{JORGE P\'ARAMOS}

\address{Instituto de Plasmas e Fus\~ao Nuclear, Instituto Superior T\'ecnico,\\Avenida Rovisco Pais 1, 1049-001 Lisboa, Portugal\\
paramos@ist.edu}

\maketitle

\begin{abstract}

We consider the feasibility of using the Galileo Navigation Satellite System to constrain possible extensions or modifications to General Relativity, by assessing the impact of the related additions to the Newtonian potential and comparing with the available observables: the relative frequency shift and the time delay of light propagation. We address the impact of deviations from General Relativity based on the parameterized Post-Newtonian parameters due to the presence of a Cosmological Constant, of a constant acceleration like the putative Pioneer anomaly, a Yukawa potential term due to massive scalar fields and a power-law potential term, which can arise from Ungravity or f(R) theories.

\end{abstract}

%\pacs{04.80.Cc, 04.25.Nx, 07.87.+v}

\section{Introduction}

The Galileo positioning system is an important step towards the improvement and development of new applications in navigation monitoring and
related topics. Its operational use of precision clocks in orbit and comparison with those on ground stations enables one to view it broadly as a timing experiment in outer space. Hence, Galileo offers a great opportunity for fundamental research in physics: together with the deployed Global Positioning System (GPS) and GLONASS systems, satellite navigation is indeed the first technological application where relativistic effects are taken into account as a regular engineering constraint on the overall design (see Refs. \cite{Ashby,Pascual,Bahder} and references therein).

As such, there are several effects arising from special and General Relativity (GR) that must be taken into account, {\it i.e.} time dilation, gravitational blueshift and the Sagnac
effect. These may yield a clock deviation of as much as $\sim 40~\mu s/day$: many
orders of magnitude above the precision of the onboard clocks considered in the mentioned navigations systems, which is of the order of $4~ns/day$, {\it i.e.} a time stability of about $5 \times 10^{-14}$.

Furthermore, the gravitational frequency shift is of the order of $V_N /c^2 \simeq 10^{-10}$ (where $V_N = GM_E /R_E$ is the Newtonian potential, $G$ is Newton's constant, $M_E \simeq 6.0 \times 10^{24}~kg$ is the Earth's mass, $R_E \simeq 6.4 \times 10^6~m$ is
its radius and $c$ is the speed of light). In the Galileo Navigation Satellite System, this correction is accounted by the receiver, while the GPS system accounts for this mismatch through an offset in the onboard clock frequency.

The Galileo navigation system offers a positioning improvement of at least one order of magnitude, from an everyday error margin of $\sim 10~m$ with the GPS system to $\sim 1~m$; a spatial accuracy of the order of $1~mm$ is possible using the planned thirty satellite constellation and carrier phase measurements, similar to the increased precision of the GPS system using averaging and modern geodetic GPS receivers. Given this, a legitimate question arises: what are the possible implications for fundamental physics that one may extract from this increased precision? 

In this study, we aim to establish bounds on the detectability of extensions and modifications to GR, by assessing the impact of the related additions to the Newtonian potential on the observables made available by the Galileo system, namely the relative frequency shift $\ep_f \equiv f/f_0 - 1$ and the light propagation time delay $\De t$. This may be done for a wide variety of already available phenomenological models (see Refs. \cite{review,Will} for updated surveys). Hence, this study should be regarded as an extension of the excellent treatments already available that deal with physical effects arising from GR \cite{Ashby,Pascual,Bahder,Wolf}, but in the context of extensions of the latter.

We assume as typical values $\ep_f = 10^{-15}$ (a precision possible using next generation Rubidium clocks) and $\De t = 10^{-12}~s$ (corresponding to a $1~mm$ precision); throughout the text, results stemming from the comparison with $\De t = 10^{-9}~s$ will also be given, so the reader may grasp the observable range obtained for spatial accuracies ranging between $1~m$ and $1~mm$.

The use of the global positioning system to ascertain the effects induced by modifications of gravity here assessed would involve the direct comparison between the received signal and its emitted time and frequency. However, one recalls that the actual frequency emitted by the satellites is not known, but may be inferred from comparison with clocks at the ground segment, assuming that they behave similarly (which appears reasonable, given the great stability of the onboard and ground clocks).

In this approach, one does not consider the increased precision due to averaging procedures or interpolation of the signals transmitted by the full constellation of satellites. A more evolved approach could resort to the redundancy introduced by the various elements of the constellation to eliminate systematic errors, correct for atmospheric and propagation effects, validate the comparison with the ground clocks, {\it etc.}. Regarding the latter point, one remarks that by comparing several received signals with the master clocks on the ground, one could lift the degeneracy between ``true'' frequency shifts and those arising from a slow, unpredictable drift away from the base rate of the emitter clocks (constrained by their $\sim 10^{-15}$ stability).

A more evolved use of the multitude of available satellites would rely on the construction of the so-called GPS (or emission) coordinates, whereas one resorts to four time coordinates, instead of one time plus three spatial ones, to define a suitable frame of reference \cite{Rovelli,Blagojevic,LachiezeRey,Coll}. These are obtained via four material bodies exchanging light signals, and are thus naturally implemented by global positioning systems. From a theoretical point of view, they may be regarded as a realistic implementation of Einstein's point coincidence timing procedure.

The realization of these GPS coordinates requires solely that one is able to measure the direction and arrival time of the emitted signal, as well as its redshift --- precisely the quantities we will use as observables in this study. In this fashion, the GPS coordinates allow one to directly read the metric components in the related frame and, via a suitable transformation, to another frame ({\it i.e.} one expressed in terms of time as measured by an observer and the usual spherical coordinates).

After this, one could then compare with the metric (in a rotating frame) arising from the approximately spherical mass distribution of the Earth, and interpret the differences as due to fundamental modifications of gravity. The accuracy and stability of the on-board clocks is reflected in the uncertainty of this transformation: again, combinations of the available satellites into sets of four could be used to obtain a group of GPS coordinates and observables that might help to eliminate systematics and other sources of error.

This paper is organized as follows: firstly, we briefly review the main relativistic effects that are accounted for in the Galileo navigation system. We then proceed and consider the possibility of measuring several corrections to the law of gravity using Galileo:

\begin{itemize}
\item Deviation from GR based on the parameterized Post-Newtonian parameters

\item Presence of a Cosmological Constant
\item Constant acceleration like the putative Pioneer anomaly
\item Yukawa addition mediated by massive scalar fields
\item Power-law addition, which can arise from Ungravity or $f(R)$ theories

\end{itemize}

Finally, conclusions are drawn and an outlook is presented.

\section{Main relativistic effects}

As stated, the purpose of this study is to assess the feasibility of the Galileo system as a probe of extensions of GR. In order to better contextualize 
this, the effects arising from this theory that are relevant for the implementation of a global navigation satellite system are briefly reviewed, according 
to discussions that can be found in Refs. \cite{Ashby,Pascual,Rovelli,Bahder}. The excellent discussion of Ref. \refcite{Ashby} on GR and the GPS system is followed, 
with the suitable adaptations and computations for the Galileo system, for comparison. The reader who is familiarized with this discussion 
may skip these introductory paragraphs and proceed directly to the new results found from Section \ref{PPN} onward.

\subsection{Frame of reference}

One begins by assuming that all time-dependent effects are of cosmological origin, and thus evolve over a timescale of order $H_0^{-1}$, where $H_0$ is Hubble's constant; hence, one may discard these as too small within the timeframe of interest and assume a static, spherically symmetric scenario. Given this, one considers the standard solution of the Einstein field equations for a non-spherically symmetric mass distribution --- which, in isotropic form, is given by the line element

\beqa ds^2 &=& \left(1 + {2V\over c^2} \right)(c~dt)^2 - {1 \over 1+ {2V
\over c^2} } \left( dr^2 + d\Om^2 \right)  \\ \nonumber & \cong & \left(1 + {2V\over c^2} \right)(c~dt)^2 -
\left(1- {2V \over c^2} \right) \left( dr^2 + d\Om^2 \right) ~~,\eeqa

\noindent where $d\Om^2 = r^2 (d\th^2 + \sin^2 \th d\phi^2)$ is the solid angle element, and $V$ is
the gravitational potential. In the unmodified GR scenario, this is simply the Newtonian potential

\beqa \label{multipole} V = V_N(r,\th,\phi) &=& - {GM_E  \over r }\bigg(1 - \sum^n_{i=2} \bigg [ J_n \left( {R_E \over r} \right)^n P_{n0} (\cos \th) + \\ \nonumber && \sum_{m=1}^n \left({R_E \over r}\right)^n \left[ C_{nm} \cos ( m \phi) + S_{nm} \sin (m \phi) \right] P_{nm} \cos(\th)  \bigg] \bigg) ~~, \eeqa

\noindent where $\th$ and $\phi$ are the latitude and longitude and $P_{nm}$ are Legendre polynomials of degree $n$ and order $m$; $J_n$ are the zonal harmonic coefficients, independent of the longitude; for $n \neq m$, the quantities $C_{nm}$ and $S_{nm}$ are the tesseral harmonic coefficients, while for $n=m$ they are dubbed sectoral harmonic coefficients. The three types of harmonic reflect the mass distribution of the Earth and its deviation from sphericity \cite{multipoles}. For very low eccentricity orbits, such as those found in global positioning systems, the tesseral and sectoral harmonics give rise to periodic perturbations only \cite{kozai}.

One now takes into account the rotation of the Earth with respect to this fixed-axis reference frame, with angular velocity $\om= 7.29 \times 10^{-5}~rad/s$; the so-called Langevin metric may be obtained by performing a coordinate shift $t' =t$, $r' = r$, $\th'=\th$ and $\phi' = \phi - \om t'$, yielding the line element

\beqa ds^2 &=& \left[1 + {2V\over c^2} - \left({\om r \sin \th \over
c}\right)^2 \right](c~dt)^2 -\\ \nonumber &&  2 \om r^2 \sin^2\th d\phi dt - \left(1 -
{2V \over c^2}\right) \left( dr^2 + d\Om^2 \right)~~,\eeqa

\noindent where, for simplicity, primes were dropped.

Clearly, a non-diagonal element appears, plus an addition to the gravitational
potential interpreted as a centrifugal contribution due to
the rotation of the reference frame; this leads to a definition of an effective
potential $\Phi = V - (\om r sin \th)^2/2$. The parameterization of the
Earth's geoid is obtained by taking the multipole expansion of $V$ up
to the desired order, calculating the equipotential lines $\Phi
=\Phi_0$ (the value of $\Phi$ at the Equator) and solving for $r(\th,\phi)$. The value adopted by the International Astronomical Union is $\Phi_0/c^2 = -6.969290134 \times 10^{-10}$.

In the above, the coordinate time $t$ is equal to the
proper time of an observer at infinity. Given the issue of ground to orbit clock synchronization, the metric should be rewritten with a time coordinate coincident with the proper time of clocks at rest on the Earth's surface.

Since the already discussed geoid provides one with an equipotential surface $\Phi =\Phi_0$, all clocks at rest with respect to it beat at the same rate. Hence, rescaling the time coordinate as $t \rightarrow (1+\Phi_0 / c^2) t$ yields 

\beqa \label{metricrescaled} ds^2 &=& \left[ 1 + {2 (\Phi-\Phi_0) \over c^2} \right] (c ~dt)^2 - \\ \nonumber && 2 \om r^2 sin^2\th d\phi dt - \left(1 - {2V\over c^2}
\right)\left( dr^2 + d\Om^2 \right)~~. \eeqa

\noindent In the above, a similar transformation of the spatial coordinates (realized in the usual scheme of GPS and Galileo) is not performed, for simplicity. Going back to an inertial, non-rotating frame, one finally writes the metric as

\beq ds^2 = \left[ 1 + {2 (V-\Phi_0) \over c^2} \right] (c ~dt)^2 -
\left(1 - {2V\over c^2} \right)\left( dr^2 + d\Om^2 \right)~~. \label{metricfinal} \eeq

\subsection{Constant and periodic clock deviation}
\label{deviation}

In this paragraph one discusses the difference in clock rates due to relativistic effects, {\it i.e.} the difference between coordinate time $t$ as measured in an Earth centered, non-rotating frame, and the proper time $\tau$ of the moving clocks onboard the satellites.

Keeping only terms of order $c^{-2}$, the proper time increment on the moving clock is approximately given by

\beq d\tau = ds / c = \left( 1 + {V-\Phi_0 \over c^2} -{v^2 \over 2
c^2} \right) dt~~, \label{dtau} \eeq

\noindent
so that, assuming an elliptic orbit with semi-major axis $a$ and, for simplicity, the Newtonian potential generated by a perfectly spherical body $V = V_N \simeq -GM_E / r$, one obtains \cite{Ashby}

\beq \label{proper} d\tau = ds / c = \left[ 1 - {3GM_E \over 2 a c^2} - {\Phi_0 \over
c^2} + {2GM_E \over c^2} \left( {1 \over a } - {1 \over r}\right)
\right] dt~~.\eeq

\noindent Hence, the constant correction terms are given by

\beq {3GM_E \over 2 a c^2} + {\Phi_0 \over c^2} = -4.7454 \times 10^{-10}~~,\eeq

\noindent for Galileo, and $-4.4647 \times 10^{-10}$, for the GPS. The difference stems from the slightly higher orbit for the former, $h_{Galileo} \approx 23.2 \times 10^3~km$ against $h_{GPS} \approx 20.2 \times 10^3~km$. As the above shows, the orbiting clock beats faster by about $41~\mu s /day$ (Galileo) and $39~\mu s /day$ (GPS).

Other residual periodic corrections are proportional to $(1/r - 1/a)$; one may directly integrate the corresponding term in \eq{proper}, yielding \cite{Ashby}

\beq  \De t_{periodic} = {2e \over c^2} \sqrt{GM_E a} \sin E = 4.4428 \times 10^{-10} e \sqrt{a} \sin E~s/\sqrt{m}~~, \eeq

\noindent where $E$ is the eccentric anomaly, defined by the transcendental equation $E - e \sin E = \sqrt{GM_E/a^3}(t-t_p)$, with $t_p$ the time of perigee passage. Assuming an almost circular orbit with eccentricity $e = 0.01$, this contribution can reach $23~ns$ (GPS) or $24~ns/day$ (Galileo, due to the slightly higher orbit) --- see Ref. \refcite{ecc} for a discussion on the long-term deviation from circular orbits.

These second-order effects arise from Eq. (\ref{proper}), via the substitution of the Newtonian expression

\beq {v^2 \over 2} = GM_E \left({1 \over r} - {1 \over 2a} \right)~~, \eeq

\noindent into Eq. (\ref{proper}); since the purpose of this study is to ascertain the effect of additional corrections to the Newtonian potential $V_N$, it is important to determine the  effect of these additions in this computation.

To do so, one considers a spherically symmetric addition $U(r)$ to the Newtonian potential, $V = V_N + U$. When a satellite is orbiting at $r=a$, the derived force is centripetal; together with the assumption energy conservation (since the additional contribution is assumed to be conservative), this yields

\beqa {v(a)^2 \over a} &=& V'_n(a) + U'(a) ~~, \\ \nonumber  {v(r)^2 \over 2} + V_N(r) + U(r) &=& {v(a)^2 \over 2} + V_N(a) + U(a) ~~. \eeqa 

\noindent  Considering for simplicity the spherically symmetric Newtonian potential $V_N = - GM_E/r$, this leads to

\beq \label{vr} {v^2 \over 2} = GM_E \left({1 \over r} - {1 \over 2a} \right) +  {U'_a a \over 2} + U_a - U(r) ~~, \eeq

\noindent with $U_a \equiv U(a)$ and $U'_a \equiv U'(a)$, for brevity.

Before replacing Eq. (\ref{vr}) into Eq. (\ref{dtau}), one should notice that the additional term $U$ will also shift the value of the effective potential at the Equator, $ \Phi_0 \rightarrow \Phi_0 + U_E$ (with $U_E \equiv U(a_1)$, where $a_1$ is the equatorial radius). Considering this, one obtains the generalization of Eq. (\ref{proper}),

\beqa \label{dtau2}d\tau = ds / c &=& \bigg[ 1 - {3GM_E \over 2ac^2}  -{\Phi_0 \over c^2} - {1 \over c^2}\left({U'_a a \over 2} + U_E - U_a\right) + \\ \nonumber && {2GM_E \over c^2} \left( {1 \over a} - {1 \over r} \right) +{2 \over c^2} \left[U(r)-U_a \right]  \bigg] dt~~. \eeqa

\noindent Solving for $dt$ and taking only $c^{-2} $ terms, one obtains

\beqa \label{corr} \De t = \int_{path}  d\tau && \bigg[ 1 + {3GM_E \over 2ac^2}  +{\Phi_0 \over c^2} + {1 \over c^2}\left({U'_a a \over 2} + U_E - U_a\right) + \\ \nonumber && {2GM_E \over c^2} \left( {1 \over r} - {1 \over a} \right) +{2 \over c^2} \left[U_a - U(r) \right]  \bigg]~~.  \eeqa

\noindent Hence, perturbing the Newtonian potential yields an additional constant rate correction $[U'_a a/2 + U_E - U_a]/c^2$, as well as a periodic term $2(U_a-U)/c^2$.

\subsection{Doppler Effect}
\label{dopplereffect} 

The transmission of a light signal between two moving bodies affects the frequency $f_S$ of the GPS signal by the usual special relativistic (first-order) Doppler effect, as well as the already mentioned second-order effect in Eq. (\ref{dtau}). Since the correction mechanism only takes the constant clock rate shift  $ (2\Phi_0 + 3GM_E) / 2 a c^2 $ into account, the 
received frequency $f_E$ will be shifted with respect to the corrected frequency $f_S $ by the linear Doppler effect, the unaccounted additional constant term and the Newtonian and perturbative periodic terms,

\beq f_E = f_S \left[1 - {U'_a a \over 2c^2} + {U_a - U_E \over c^2} + {2GM_E \over c^2} \left({1 \over a} - {1 \over r}\right) + {2(U-U_a ) \over c^2}  \right]{c - \vec{n} \cdot \vec{u} \over c - \vec{n} \cdot \vec{v}} ~~, \label{doppler} \eeq 

\noindent where $\vec{v}$ is the velocity of the satellite, $\vec{u} = \vec{R_E} \times \vec{\om}$ the rotation velocity of the geoid and $\vec{n}$ a unit vector in the direction of propagation (as seen in a local inertial frame).

As seen in Eq. (\ref{vr}), if an addition to the Newtonian potential is regarded as a perturbation, then the velocity of a satellite orbiting at a radius $r$ is given by

\beq {v \over c} \approx \sqrt{{GM_E \over c^2} \left({2 \over r} - {1 \over a}\right)} + {1 \over c^2}\left[U_a - U + {U'_a a \over 2} \right] \sqrt{a c^2\over GM_E} ~~.\label{vrapprox}\eeq

\begin{figure} 
\center
\epsfxsize=\columnwidth \epsffile{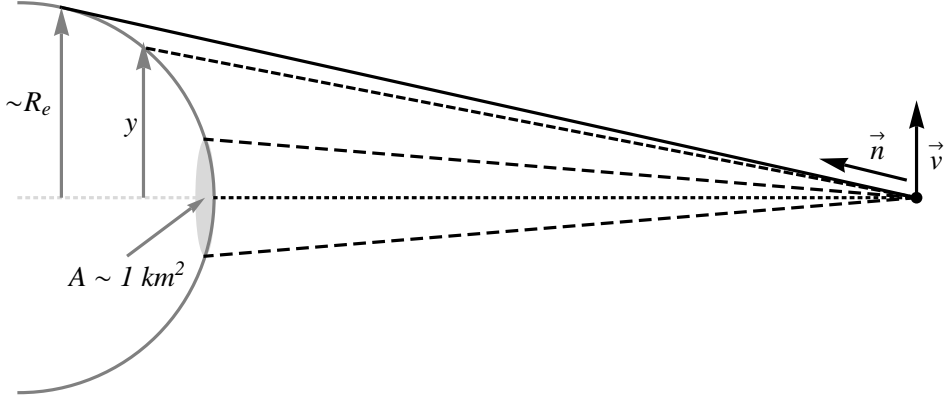}
\caption{Signal path geometry: the gray area  (not to scale) gives the $\sim1~km^2$ area where the linear Doppler term is negligible, $\de \ll 1$. ``Nose down'', $y=0$, and grazing transmissions, $y \sim R_E$, are also depicted (dotted and full line, respectively).}
\label{grazing}

\end{figure}

Following Fig. (\ref{grazing}), the deviation of the Doppler factor at the end of Eq. (\ref{doppler}) is given by

 \beq {\vec{n} \cdot \vec{v} \over c} = {y \over \sqrt{y^2 + (R_E+h)^2}} {v \over c} \approx {y \over R_E + h} {v \over c} \sim 3 \times 10^{-6}{y \over R_E} ~~. \eeq

\noindent The maximum value is obtained for a grazing signal $y \sim R_E$ (since $h > R_E$); the perturbation $U$ will of course modify its magnitude, due to the shift in the velocity.

Given this, comparison of Eq. (\ref{vrapprox}) with (\ref{doppler}) indicates that the linear Doppler shift affects the contributions of the constant clock rate term $-U'_a a /2c^2$ and the periodic term $(U-U_a)/c^2$. At leading order, the relative frequency shift induced by $U$, dubbed $\ep_f$, is

\beq \label{relfreq0} \ep_f = - \left[ 1 - \de(y)\right] {U'_a a \over 2c^2} + {U_a - U_E \over c^2} + \left[1 - {\de(y) \over 2} \right]{2(U-U_a ) \over c^2}   \eeq

\noindent with the term arising from the linear Doppler shift being defined as 

\beq \de(y) \equiv {y \over R_E + h} \sqrt{(R_E + h) c^2 \over GM_E} \eeq

\noindent setting $a = R_E + h$.

Clearly, the competition between the second-order effect embodied in the metric Eq. (\ref{corr}) and the linear effect of the kinematic Doppler factor depends upon the angle between the velocity of a satellite and the direction of propagation, {\it i.e.} the ratio $y/R_E$. one may ascertain the values for which the linear effect cancel the second-order one in Eq. (\ref{relfreq0}):

\beq \label{defde} \de(y)=1 \rightarrow  {y \over R_E} = \left(1 + {h\over R_E }\right) \sqrt{ GM_E \over (R_E + h) c^2} \sim 5 \times 10^{-5}  \eeq

\noindent and $\de(y)=2 \rightarrow y \sim 10^{-4} R_E$. Hence, one concludes that only for an almost radial transmission (``nose down'', with a spread of less than $1~km$) is the linear term smaller than the second-order one. 

One thus assumes the scenario of a grazing transmission $y = R_E$, where the linear term is enhanced and the possibility of detecting an addition to the gravitational potential is boosted. A more realistic choice would involve a shorter transmission path $y < R_E$, thus avoiding the issue of transversing the thick atmospheric layers and deal with the ensuing signal distortion. Also, the relation between the different orbital planes adopted by each global navigation satellite constellations and the rotation axis of the Earth would have to be properly accounted for, namely through a careful evaluation of the Sagnac effect for each case.

With the above in mind, one writes 

\beqa \label{relfreq} \ep_f &=& \ep_c + \ep_p ~~, \\ \nonumber \ep_c &=& {U_a - U_E \over c^2} +  \de(R_E) {U'_a a \over 2c^2} ~~~~,~~~~\ep_p = \de(R_E){U_a-U \over c^2}   \eeqa

\noindent From Eq. (\ref{defde}), one sees that the parameter $\de(R_E)$ decreases with $\sqrt{h}$; since $h_{Galileo} \approx 23.2 \times 10^3~km$ against $h_{GPS} \approx 20.2 \times 10^3~km$, one has $\de(R_E)_{Galileo} = 1.7 \times 10^4$ versus $\de(R_E)_{GPS} = 1.8 \times 10^4$. Thus, the GPS global navigation system yields a marginally higher frequency shift (a $5\%$ increase of the affected terms), although insufficient against the argued advantage of using the European system increased precision to detect it.

Finally, notice that the constant frequency shift $\ep_c$ in \eq{relfreq} differs significantly from the naive expectation that would arise if one straightforwardly considered the Schwarzschild metric and simply computed the gravitational frequency shift due to $U$ (setting $V_N = 0 $),

\beq \ep_f = \sqrt{g_{00~Earth} \over g_{00~Sat}} - 1 = \sqrt{1 - 2U_E/c^2 \over 1-2U_a/c^2} - 1 \simeq  {U_a - U_E \over c^2} ~~.\eeq

\noindent The difference is not a minor one: the dominance of the linear Doppler term implies that the induced frequency shift can be several orders of magnitude higher than the above would anticipate.

In the subsequent sections, one shall use the characteristic values for this constellation, with the notation $\de = \de(R_E)_{Galileo}$, for brevity; also, one approximates the equatorial radius $a_1$ by the mean Earth radius $R_E$, and the semi-major axis as $a = R_E + h$, and computes the amplitude of the periodic clock rate term, given by

\beq \ep_p = \de(R_E){U_a-U_b \over c^2} \eeq

\noindent where $U_b \equiv U(b) $ and $b = a \sqrt{1-e^2}$ is the semi-minor axis.

\subsection{Shapiro time delay and the Sagnac effect}

The Shapiro time delay is a second order relativistic effect affecting the propagation of light \cite{Ashby}, given by

\beq \De t_{delay} = {\Phi_0 l \over c^3} + {2GM_E \over c^3}~\ln
\left(1 + {l \over R_E} \right) \simeq 6.67 \times 10^{-2}~ns~~, \label{tdelay} \eeq

\noindent a result obtained after integrating over a radial path of proper length $l$ and considering only the effect of the Newtonian potential. In the following sections the additional contribution stemming from the specific perturbative potentials $U$ will be addressed; for simplicity, one uses $l =  h$, thus assuming a ``nose-down'' propagation path.

The Sagnac effect is yet another relativistic contribution, which reflects the rotation of the Earth and the consequent difference between the gravitational potential $V$ of a non-rotational frame and the effective potential $\Phi$ of a rotating one; from the standpoint of an observer in an inertial ({\it i.e.} non-rotating) frame, it reflects the added propagation delay due to the rotation of the receiver as light travels towards it. From Eq. (\ref{metricrescaled}), one gets the additional time delay

\beq \De t_{Sagnac} = { \om \over c^2} \int_{path} r^2 ~d\phi = {2 \om
\over c^2} \int_{path} dA_z ~~, \eeq

\noindent where the integral gives the equatorial projection of the area swept out by the light signal in its travel between the satellite transmitter and the rotating ground receiver. Its magnitude it highly dependent on the assumed path: given the typical configuration of ground segment stations and the location of the satellites, it varies between $240 ~ns$ and $350~ ns$ \cite{AshbySagnac}.

As already shown above, a global positioning system is affected by a frequency shift of order $ 10^{-10}$ and a cumulative propagation time delay of the order of $0.1~\mu s$ (mainly due to the Sagnac effect). The following sections aim to compute other corrections that should be taken into consideration when calculating the clock synchronization of ground and onboard clocks, and compare the obtained results with the precision of $10^{-15}$ and the time accuracy of Galileo, of order $10^{-12}~s$.

\section{Post-Newtonian effects}
\label{PPN}

Before venturing into more speculative and hypothetical effects, arising from extensions to GR, it is natural to first tackle the possibility of measuring Post-Newtonian effects with the Galileo positioning system; these are naturally much smaller than the previously considered, of higher order in the Newtonian potential, $V_N /c^2 \approx GM_E/(R_Ec^2) \sim 10^{-10}$ (again considering a purely spherical body).

Post-Newtonian effects are suitably addressed by resorting to the so-called parameterized Post-Newtonian (PPN) formalism, which allows one to describe higher-order effects induced by metric extensions and alternatives to GR. For simplicity, one focuses only on the $\be$ and $\ga$ PPN parameters, thus writing the related PPN metric \cite{Will,Klioner,PPN2,Kopeikin} as

\beq ds^2 = - \left[ 1 + {2V \over c^2} + 2 \be \left({ V \over c^2}\right)^2 \right]~(c~dt)^2 + \left( 1 + 2 \ga {V \over c^2} \right)~\left( dr^2 + d\Om^2 \right) ~~. \label{emetric} \eeq

\noindent The $\be$ parameter measures the amount of non-linearity affecting the superposition law for gravity, while $\ga$ is related to the spatial curvature per unit mass.

Although not evoked here, the full PPN metric includes a total of ten PPN parameters, which characterize the underlying fundamental theory and its possible consequences; these may include a violation of momentum conservation, existence of a privileged reference frame, amongst others deviations from GR. The PPN formalism is defined so that General Relativity is parameterized by $\be=\ga=1$, while all remaining parameters vanish; measurements of the Nordtvedt effect yield $|\be -1 | \leq (2-3) \times 10^{-4}$, while Cassini radiometry indicates that  $\ga -1 = (2.1 \pm 2.3) \times 10^{-5}$ \cite{beta,gamma}.

Also, one recalls note that the gravitational potential $V$ appearing above is, in principle, due to the matter and energy distribution of the whole Universe, through

\beq V(\vec{x}) = \int {\rho(\vec{x}',t) \over  |\vec{x} - \vec{x}'|} d^3 x' ~~, \eeq

\noindent with similar definitions for the metric potentials involved in the full PPN metric (with the aforementioned ten parameters), which include not only the density distribution, but also the velocity field of the source matter (see Ref. \refcite{Will} for a full discussion).

Notice that the PPN potentials do not depend on the PPN parameters, only on the matter density, pressure and velocity: the latter are coefficients affecting the expansion of the metric elements in terms of these quantities, as Eq. (\ref{emetric}) shows. One may neglect the contributions of distant sources to these potentials, independently of possible deviation of the PPN parameters from their GR values (experimentally constrained to be very small); in other words, no PPN parameter directly enforces a Machian view of gravitation, in which distant sources would directly induce strong local effects on the gravitational field.

With the above in mind, one writes the acceleration in the weak-field limit,

\beq \vec{a} = \Ga_{00}^r \approx -{1 \over 2} g^{rr} g'_{00} \approx  -{GM_E \over r^2} \left[ 1 + 2(\ga-\be){GM_E \over c^2 r} \right] ~~, \eeq

\noindent where the prime denotes differentiation with respect to the radial coordinate.

The experimental bounds discussed above indicate that the difference $(\ga-\be)$ should be of order $\lesssim 10^{-4}$; since $GM_E /c^2 R \sim 10^{-10}$, one concludes that PPN corrections to the acceleration are much too small to be detected. Likewise, similar expressions may be derived for the time delay and frequency shift, showing that the post-Newtonian relative corrections are indeed proportional to $V_N/c^2 = GM_E/(R_Ec^2) \sim 10^{-10}$. Given the already considered gravitational time delay and frequency shift are of the order $\De_t \sim 10^{-9}~s$ and $\ep_f \sim 10^{-10} $, respectively, comparison with the available precision of Galileo and GPS systems makes it clear that Post-Newtonian effects signaling deviations from GR are much below the observation threshold.

%%%%%%%%%%%%

\subsection{Strong Equivalence Principle}

There is another impact of deviations from GR within the PPN formalism that must be considered. Indeed, while distant bodies do not impose a strong acceleration on the individual satellites, the effect of the small gradient between their effect on the latter, and on the Earth itself --- that is, a tidal acceleration --- is boosted if the Strong Equivalence Principle (SEP) is not valid \cite{review}. This difference would be perceived on the ground as an acceleration of the satellite with respect to the geocentric frame, and could in principle be relevant. Thus, the purpose of this section is to ensure that such an effect is under control in the context of the approach of this work.

The SEP implies that the gravitational field properties reflect the gravitational energy of the bodies themselves, and provides an assumption about the nonlinear properties of gravity. GR assumes that this principle is exact, but alternative metric theories of gravity, such as those involving scalar fields, and other extensions, typically violate the SEP \cite{Nordtvedt,Nordtvedt2,Nordtvedt3,Nordtvedt4}.

In the framework of the PPN formalism (and assuming fully conservative, Lorentz-invariant theories, for simplicity \cite{Will,Kopeikin}), the validity of the SEP translates into the vanishing combination $\eta = 4\be - \ga - 3 = 0$ --- trivially satisfied for GR, where $\be = \ga = 1$. If $ \eta \neq 0$, the acceleration of a body with inertial and gravitational mass $m$ and $m_g$, respectively, is given in the Newtonian limit by

\beq \vec{a} = {m_g \over m} \nabla V~~. \eeq

\noindent The ratio between gravitational and inertial masses is given by

\beq {m_g \over m} = 1 - \eta {E \over m c^2} ~~, \eeq

\noindent with the (negative of) the gravitational self-energy $E$ given by

\beq E = {G \over 2} \int_{body} d^3 x d^3 y {\rho(\vec{x}) \rho(\vec{y}) \over || \vec{x}-\vec{y}||} ~~. \label{selfenergy} \eeq

Let us now consider two bodies with inertial mass $m_1$, $m_2$ in the gravitational field of an external object with mass $M$ at a distance $L$; the bodies are separated by a small distance $l$ when compared to $L$. For simplicity, one considers only the case where the three bodies are aligned, whereas the tidal acceleration, given by the difference between individual gravitational pulls, attains its higher value:

\beqa \label{tidal} a_T &=& a_2 - a_1 = G M \left[ {m_{g1} \over m_1} { 1 \over (L-l)^2} -{m_{g2} \over m_2} {1 \over L^2} \right] \\ \nonumber & \approx& {GM \over L^2} \left[ {m_{g1} \over m_1} \left( 1 +{2l \over L} \right) - {m_{g2} \over m_2} \right] \\ \nonumber &\approx& {GM \over L^2} \left[ {2 l \over L}  + { \eta \over c^2} \left( {E_2 \over m_2 } -  {E_1 \over m_1 }  \right) \right] ~~,  \eeqa

\noindent using $l \ll L$.

The term $a_{TN} = 2GM l /L^3$ is the Newtonian tidal acceleration. Considering the Earth and one satellite, $m_1 = M_E$ and $m_2 = m \approx 700~kg$, separated by a distance $l = R_E + h \approx 3 \times 10^7~m $, one may compute the ratio between this term and the gravitational pull of the Earth on a satellite, $a_E = GM_E / l^2$:

\beq {a_{TN} \over a_E} = 2 {M \over M_E} \left({l \over L}\right)^3~~, \eeq

\noindent One can obtain this ratio for the relevant bodies in the Solar System: the Sun ($M = 3.3 \times 10^5 M_E$ and $L = 1~AU $) yields $a_{TN}/a_E \sim 10^{-6} $; Jupiter at its closest point from Earth ($M = 318 M_E$ and $L = 4.2~AU $) produces $a_{TN}/a_E \sim 10^{-10}$; finally, the Moon ($M=1.23 \times 10^{-2} M_E$ and $L = 3.84 \times 10^8~m$) leads to $a_{TN}/a_E \sim 10^{-5}$. Thus, one concludes that the tidal forces induced by external bodies may be disregarded if the SEP holds and $\eta =0$.

This said, there is the possibility that SEP violating effects might induce additional, relevant tidal effects, as expressed by the second term of Eq. (\ref{tidal}),

\beq a_{T\eta} = \eta {GM \over L^2 c^2} \left({E_E \over M_E } - { E_S \over m}\right)~~,\eeq  

\noindent where $E_E = 4.6 \times 10^{-10} M_E c^2$ is the gravitational self-energy of the Earth \cite{Will}, and $E_S$ the corresponding quantity for a navigation satellite. Indeed, one notes that the Newtonian tidal acceleration drops for increasingly distant external bodies (as $L^3$), while the term arising from SEP violation falls only quadratically, thus eventually surpassing the former.

The extreme smallness of gravitational self-energy for laboratory-sized objects\footnote{It is of the same order of magnitude as a sphere with radius $R \sim 1~m$ and uniform density, yielding $E_S / mc^2 = (3 / 5) (Gm / Rc^2) \sim 10^{-25}$.} implies one has to consider astrophysical scales in order to test the SEP. Currently, the Earth---Moon---Sun system provides the best scenario for testing this principle, manifested in the so-called Nordtvedt effect \cite{Nordtvedt}. Lunar laser ranging experiments \cite{llr} yield the constraint $\eta = (4.4 \pm 4.5) \times 10^{-4} $ \cite{progress}. 

One can easily check that no Solar System body can lead to a SEP-violation induced term larger than the Newtonian tidal acceleration $a_{TN}$. Indeed, the ratio between the two is

\beq {a_{T\eta} \over a_{TN}} \approx \eta {E_E \over M_E c^2}{L \over 2l} < 2 \times 10^{-13} {L \over l}~~.\eeq

\noindent Taking the highest allowed value yields the upper bound $\eta < 8.9 \times 10^{-4}$.

The SEP violation induced term is important only if the above ratio is much larger than unity, {\it i.e.} when the distance between the external body and the Earth--satellite system is $L \gg 5 \times 10^{12} l \sim 5~kpc$! Thus, one concludes that only objects at a galactic distance might yield relevant effects.

The above result only states that for galactic ranges, the SEP violating tidal term is larger than the Newtonian one, $a_{T\eta} > a_{TN}$, but not necessarily measurable. To assess this, one considers the ratio between the former and the gravitational pull of the Earth $a_E$,

\beq {a_{T\eta} \over a_E} \approx \eta {M \over M_E} \left({l \over L}\right)^2 {E_E \over M_E c^2} < 4 \times 10^{-13} {M \over M_E} \left({l \over L}\right)^2~~,\eeq

\noindent again considering  the highest allowed value for $\eta$. Considering the Milky Way mass to be concentrated at its core, $M = M_{MW} = 2.3 \times 10^{17}M_E$ and $L = 7.62~kpc$, one obtains $a_{T\eta} /a_E < 1.6 \times 10^{-15}$; Virgo, the closest cluster to Earth, at a distance $L = 16.5~Mpc$ and with a mass $M = 4.0 \times 10^{20} M_E$, yields $a_{T\eta} /a_E < 5.7 \times 10^{-25}$.

One now recaps the results of this section: although external bodies at a sufficiently large distance (at the galactic range) may induce SEP violating tidal effects larger than their Newtonian counterpart, these are clearly much below the gravitational pull of the Earth: one finds that the external bodies are not sufficiently massive to compensate for their extreme remoteness. As a result, one can safely state that SEP violating effects can be disregarded in the context of the present study; conversely, one does not have to assume that this fundamental tenet of GR is obeyed, {\it i.e.} the discussion presented before concerning the PPN parameters is not constrained by the condition $\eta = 4\be -\ga -3 =0$.

%%%%%%%%%%%%%

\section{Detection of the cosmological constant}

According to the latest observations, the Universe is currently ongoing a period of accelerated expansion; although several proposals exist to account for this acceleration, the simplest explanation resorts to a cosmological constant $\La \sim 10^{-35}~s^{-2}$, which acts as a fluid with negative pressure (see Ref. \refcite{Lambda} and references therein). The local effect of this component may be evaluated by matching the outer Friedmann-Robertson-Walker metric with a static, symmetric solution given by Birkhoff's theorem; this yields the Schwarzschild-de Sitter metric\footnote{As a side note, one remarks that the identification of the time coordinate $t$ of the current form as the proper time of an observer at rest at infinity breaks down, due to the collapse of the Schwarzschild ``bubble'' at a finite distance, where it matches the exterior FRW metric \cite{bubble}.}, with a line element \cite{SdS,Claus},

\beq ds^2 = -\left(1+{2 (V_N + U_\La)\over c^2 } \right) (c~dt)^2 + \left[ 1+{2 (V_N + U_\La) \over c^2 } \right]^{-1} dr^2 + d\Om^2~~.\eeq

\noindent in anisotropic form; the presence of a cosmological constant produces an additional potential term $U_\La = -\La r^2/6$ (with $\La = 3\Om_\La H_0^2$ and $\Om_\La = 0.7$).

In order to compare with the metric Eq. (\ref{metricfinal}), a coordinate change to an isotropic, co-rotating frame of reference should be performed. However, given the weak field regime, this would only amount to a small correction to the effect of this additional potential, which is in itself a perturbation to the Newtonian one.

\subsection{Constant frequency shift}

Following \eq{relfreq}, the constant frequency shift induced by this extra potential contribution is easily obtained,

\beq \ep_c \simeq -\de{\La \over 6c^2}(R_E+h)^2 \sim -10^{-34}~~, \eeq

\noindent clearly much below the $\ep_f = 10^{-15} $ precision available.

\subsection{Periodic frequency shift}
\label{periodicCC}

The maximum periodic frequency shift is given by

\beq \ep_p = -\de {\La \over 6c^2}(R_E+h)^2 e^2 = \ep_c e^2 \ll \ep_c \ll \ep_f~~, \eeq

\noindent and so is impossible to detect.

\subsection{Propagation time delay}

Similarly, the additional propagational time delay is given by

\beq \De t_{delay}={1 \over c^3} \int_{R_E}^{R_E+h} U_\La(r)~dr~~, \eeq

\noindent so that the presence of a cosmological constant results in a further contribution of

\beq \De t_\La = {1 \over c^3}  \int_{R_E}^{a} {\La r^2 \over 6}~dr =  {\La \over 18c^3} h \left[ (3R_E(R_E+h) + h^2\right] \sim 10^{-39}~s~~, \eeq

\noindent also many orders of magnitude below the optimum time resolution of $10^{-12}~s$. Therefore, one concludes that the cosmological constant is completely undetectable by the Galileo positioning system (as indicated by an analytical study in Ref. \refcite{Eva}).

\section{Detection of anomalous, constant acceleration}

Although not usually considered, the presence of an anomalous, constant, acceleration affecting the free-fall of bodies could model effects arising from some fundamental theory of gravity, perhaps hinting at the existence of a fundamental threshold between known dynamics and yet undetected exotic physics.

One widely discussed example is the so-called Modified Newtonian Dynamics (MOND) model \cite{Milgrom,Bekenstein,MOND}, which features a departure from the classical Poisson equation at a characteristic acceleration scale of about $10^{-10}~m/s^2$, and aims to explain the puzzle of the galaxy rotation curves without evoking any dark matter component.

From the experimental viewpoint, a constant acceleration  $a_P=(8.74 \pm 1.33) \times 10^{-10}~m/s^2$ is reported to affect the Pioneer 10/11 probes \cite{pioneer1,pioneer2,pioneer3}; its origin, either due to an incomplete engineering analysis (see {\it e.g.} Ref. \refcite{thermal1,thermal5,thermal6,thermal2,thermal3,thermal4}) or stemming from yet undiscovered fundamental physics \cite{Paramos,Reynaud,Moffat}, has been dubbed the Pioneer anomaly.

\subsection{Constant frequency shift}

An anomalous, constant acceleration, $a_C$, would imply an addition to the gravitational potential of the form $U_C = a_Cr$; following the procedure depicted in the previous section, one calculates the related constant frequency shift as

\beq \ep_C \simeq \de{a_C \over 2c^2}(R_E+h) ~~, \eeq

\noindent and, comparing with the frequency stability $\ep_f = 10^{-15}$, one obtains an upper bound for a detectable constant acceleration $a_C$,

\beq a_C \gtrsim {2 \ep_f c^2 \over \de(R_E+h)} = 3.5 \times 10^{-10}~m/s^2~~. \eeq

This range allows for a narrow fit of the reported Pioneer anomalous acceleration $a_P$, which yields $\ep_C = 2.5 \times 10^{-15}$; notice that previous computations \cite{toulouse,padova} without the linear Doppler shift yielded a completely negligible $\ep_C$, which shows the enhancing effect of the former.

Since a more evolved treatment is not pursued here (including realistic locations of ground segment stations, atmospheric correction, {\it etc.}) and this unobserved correction is at the edge of detectability, one cannot regard this finding as a major point against of the existence of the Pioneer anomaly --- which, given the conclusions of Refs. \refcite{thermal1,thermal5,thermal6,thermal2,thermal3,thermal4}, now presents an unaccounted acceleration of about only one third its originally reported value $a_P$.

\subsection{Periodic frequency shift}

Similarly to the result of paragraph \ref{periodicCC}, the maximum periodic frequency shift is given by

\beq \ep_p \simeq \de{a_Ce^2\over 2c^2}(R_E+h) = \ep_c e^2 \ll \ep_c~~, \eeq

\noindent and is therefore of no use in further constraining the allowed range of the anomalous acceleration $a_C$.

\subsection{Propagation time delay}

The propagational time delay due to this extra potential addition is given by

\beq \De t_C = {1 \over c}  \int_{R_E}^{R_E+h} {ar\over c^2}~dr = {a \over 2 c^3}h(2R_E+h)~~. \eeq

\noindent By the same token, comparison with a time accuracy of $10^{-12}~s$ indicates that only accelerations $a \gtrsim 0.1~m/s^2 $ are measurable using time delay.

\section{Detection of Yukawa potential}

A Yukawa potential is one of the more discussed modifications to the law of gravity, as it may arise from scalar-vector-tensor field ``fifth force'' models, where its characteristic range $\la$ is related to the mass $m$ of the scalar or vector field, $\la \propto m^{-1}$ \cite{review}.

For the case of exchange of a scalar particle, the full potential is given by

\beq V(r) = -{G_\infty M_E \over r} \left(1 + \al e^{- r /\la} \right)~~, \eeq

\noindent where $\al < 1$ is the strength of the perturbation and $G_\infty$ the gravitational coupling for $r \rightarrow \infty$; the latter redefines Newton's constant $ G $ through the relation $G = G_\infty (1+\al)$; this full potential may be separated into a Newtonian-like potential and an extra potential 

\beq U_Y (r) = -{\al \over 1+\al} {G M_E \over r}e^{-r/\la}~~.\label{yukawapot} \eeq

\begin{figure} 

\epsfxsize=\columnwidth \epsffile{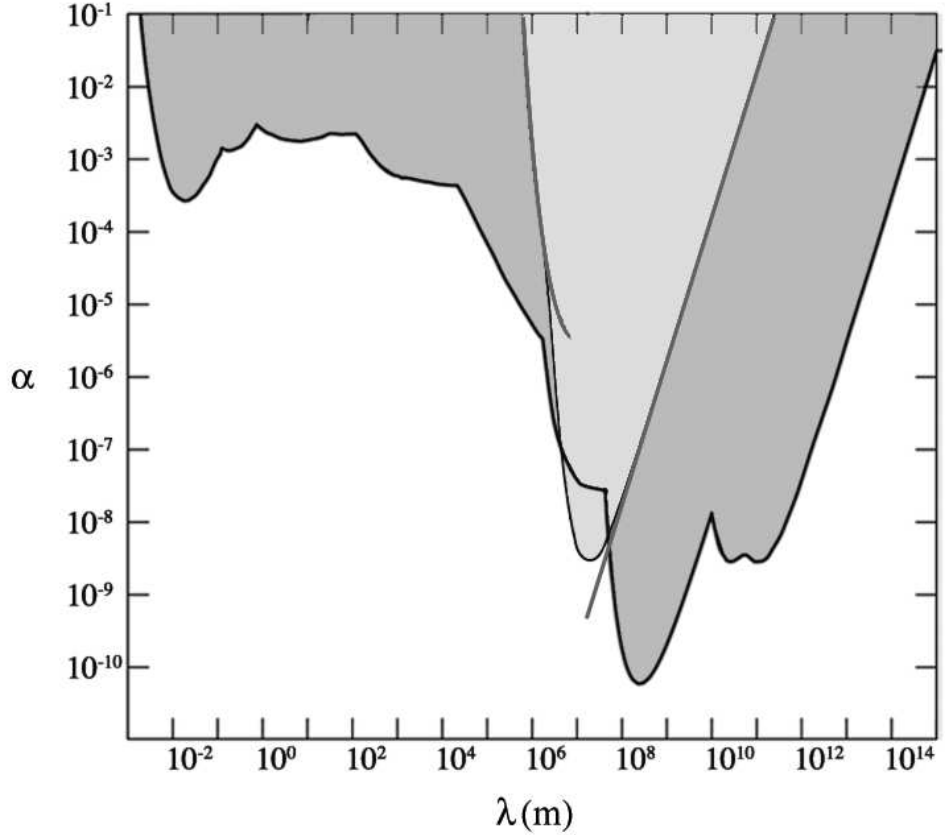}
\caption{Exclusion plot (shaded region) for a Yukawa-type additional force with strength $\al$ and range $\la$: dark gray corresponds to the available bounds $^{46}$ and light gray gives the superimposed excluded range obtained for a relative frequency precision $\ep_f =10^{-15}$; approximations in the short- and long-range regimes are depicted as gray lines.}
\label{yukawa}

\end{figure}

By conjugating several constraints arising from different setups covering a wide range of distances (from near-millimeter tests to planetary experiments), stringent bounds have been obtained \cite{fischbach} for the allowed region of parameter space $\al,~\la$, as may be seen in Fig. \ref{yukawa}. 

One now computes the constant relative frequency shift,

\beq \ep_c = {\al GM_E \over (R_E+h)c^2 } e^{-(R_E+h)/\la}\left[ \left(1 + {h \over R_E}\right) e^{h/\la} - 1 + {\de \over 2} \left( {R_E + h \over \la} + 1 \right) \right]~~. \label{epcy} \eeq

\noindent Since the magnitude of $\la$ is not known, terms in $\de$ may not dominate this expression. To ascertain this, one must consider the short- and long-range regimes separately.

The maximum periodic term is given by

\beq \label{eppy} \ep_p \simeq \de{\al GM_E \over (R_E+h)c^2}e^{-(R_E+h)/\la} \left[ e^{(R_E+h)e^2/2\la} \left(1+{e^2 \over 2}\right) - 1 \right]  \eeq

\subsection{Short-range ``fifth force''}

\subsubsection{Constant frequency shift}

One first considers that the additional Yukawa interaction is short-ranged, $\la \ll h,~R_E$. In the regime $\la < R_E$, the Yukawa potential term \eq{yukawapot} is further affected by a form factor that reflects the extended mass distribution of the source (the Earth) and the exchange of massive particles of mass $m = \la^{-1}$ only within a spherical shell of limited width.

This form factor serves to further suppress the already exponentially small effect: the constant frequency shift \eq{epcy} is approximated by

\beq \ep_c = {\al GM_E \over R_E c^2 } e^{-R_E/\la} ~~, \eeq

\noindent and the Doppler linear term is negligible for

\beq 2 \la e^{h/\la}  > \de R_E  \rightarrow \la < 0.36 R_E ~~, \eeq

Thus, one does not pursue the computation of the effects of a short-range Yukawa potential. Fig. (\ref{yukawa}) depicts the numerical computation of the constraint arising from the constant frequency shift with no form factor included: one concludes that no bound on the parameter space $(\la,\al)$ can be obtained in this regime, even with this unrealistic assumption.

\subsection{Long-range ``fifth force''}

\subsubsection{Constant frequency shift}

One now examines the opposite assertion concerning the characteristic lengthscale $\la$, and instead assumes a long range fifth force, $\la \gg h, R_E$. The exponential terms may be expanded to first order in $r/\la$ and the additional frequency shift \eq{epcy} is dominated by the linear Doppler shift term\footnote{As noted before, a previous computation \cite{toulouse,padova} without this term yielded a value for $\ep_c$ smaller by a factor $\de \sim 10^4$, again displaying the relevance of this effect.},

\beq \ep_c \simeq -\de {\al GM_E \over 4c^2}{R_E+h \over \la^2} ~~. \label{epclong} \eeq

\noindent Notice that the zeroth-order term arising from the expansion of the potential \eq{yukawapot} is absorbed into the constant $G$.

For this effect to pass unnoticed at a $\ep_f = 10^{-15}$ precision, one must have

\beq \log \al \ll \log\left({4c^2 (R_E+h)\over \de GM_E}\right) + 2\log\left({\la \over R_E+h}\right) + \log \ep_f \approx -8.81 + 2\log\left({\la \over R_E+h}\right)~~. \eeq

\noindent As can be seen in Fig. (\ref{yukawa}), this leads to a new excluded region in the $(\al,\la)$ parameter space. 

\subsubsection{Periodic frequency shift}

As in the previous paragraph, one may obtain the periodic clock rate shift \eq{eppy},

\beq \label{eppylong} \ep_p \simeq -\de{\al GM_E \over 2c^2}{R_E + h  \over \la^2} e^2 = 2 \ep_c e^2 \ll \ep_c~~,  \eeq

\noindent clearly below the constant contribution. 

\subsubsection{Propagation time delay}

The additional propagation time delay reads

\beq \De t_Y \simeq -{GM_E \al \over c^3} {h \over \la}~~.\eeq

\noindent If the effect is undetected at a level $\De t \sim 10^{-12}~s$, one obtains

\beq \al  < {c^3 \De t \over GM_E}{\la \over h} \approx 2.9 \times 10^{-9}\left( {\la \over 1~m} \right)~~. \eeq

\noindent For a lower bound of $\la \approx 10^8~m$ (only one order of magnitude above $R_E, h$), the result $\al \lesssim 0.1 $ is found, which does not advance the already excluded region of the parameter space $(\la,~\al)$ (see Fig. \ref{yukawa}).

As Fig. \ref{yukawa} shows, if clock rate shifts induced by a long-range Yukawa potential are undetected at the considered precision, then the excluded region already probed in the $\la = 10^7-10^8~m$ domain becomes even deeper. This is a novel result stemming from this approach.

\section{Detection of a power-law addition to the Newtonian potential}

One finally approaches the possibility of additions to the gravitational potential of the form

\beq U_P = -{G M_E \over r} \left( { R \over r} \right)^n~~, \label{power} \eeq

\noindent where $n \neq -1$ is a (possibly non-integer) exponent and $R$ is a characteristic length scale arising from the underlying physical theory.

Phenomenologically, such a modification of the law of gravity is an interesting alternative to the more usual Yukawa parameterization, and allows one to investigate a wider range of extensions and modifications of GR.

Such an addition can also be theoretically motivated: it arises from power-law induced effects at astrophysical scales due to the so-called Ungravity inspired scenario, which at short range involves the exchange of spin-2 unparticles of a putative scale invariant ``hidden'' sector within the Standard Model \cite{Georgi,GN}. Bounds for these ``Ungravity''-inspired corrections can be obtained from stellar stability considerations \cite{BPS09}, cosmological nucleosynthesis \cite{BS09} and the gravitational quantum well \cite{GQW}. If a power-law addition is related to Ungravity, the exponent $n$ follows from the scaling dimension of the unparticle operators $d_U$, through $n = 2d_U -2$; the lengthscale $R$ reflects the energy scale of the unparticle interactions, the mass of exchange particles and the type of propagator involved.

Other possible power-law additions to the Newtonian potential may arise from $f(R)$ theories of gravity \cite{fRa,fRb,fRc}, which generalize the Einstein-Hilbert action by considering a non-trivial scalar curvature term and/or a non-minimal coupling of geometry with matter \cite{f2}): the extra contribution arising from these models have been obtained in an astrophysical context, when addressing the puzzle of the galaxy rotation curves \cite{capo,galaxy}, as well as in a cosmological one, when applied to the issue of the accelerated expansion of the Universe \cite{cosmo}.

From Eq. (\ref{power}), one sees that the Newtonian potential $\Phi_N$ is recovered by setting $R = 0$ (for positive $n$) or 
$R \rightarrow \infty$ (for negative $n$). The limit $n \rightarrow 0$ is ill-defined, since the additional term $V_P$ does not vanish, but is instead equal to the Newtonian potential, $V_P = \Phi_N$: hence, one should rewrite the gravitational constant in terms of an effective coupling, leading to the full potential

\beq V = - {G_P M_E \over r} \left[ 1 +  \left({ R \over r} \right)^n \right]~~,\eeq

\noindent with 

\beq G_P = {G \over 1 + \left({R \over R_0}\right)^n} ~~,\eeq

\noindent where $R_0$ signals the distance at which the full gravitational potential matches the Newtonian one, $\Phi(R_0) = \Phi_N (R_0)$.

This additional length scale $R_0$ should be an integration constant, obtained after solving the full field equations that lie behind the considered power-law correction. For simplicity, one assumes that $(R /R_0)^n \ll 1$, so that this term may be safely discarded --- at the cost of neglecting the regime $n \rightarrow 0$. With this in mind, the following paragraphs use $G_P = G$ freely; one remarks that this approach is complementary to that considered in Ref. \refcite{BPS09}.

\subsection{Constant frequency shift}

As before, the constant relative frequency shift $\ep_c$ of an emitted signal is given by \eq{relfreq}, which is dominated by the linear Doppler shift,

\beqa \label{epcp} \ep_c \simeq (1+n)\de {G M_E \over 2(R_E+h)c^2} \left({R_E \over R_E+h}\xi\right)^n  ~~, \eeqa

\noindent where the dimensionless ratio $\xi \equiv R/R_E$ is defined. Comparison with a precision $\ep_f \sim 10^{-15}$ yields the constraint

\beq |1+n|\left({R_E \over R_E+h}\xi\right)^n \ll {2(R_E+h)c^2 \over GM_E} {\ep_f \over \de}~~, \eeq

\noindent which gives an upper or lower bound on $\xi$, depending on the sign of $n$. Since the {\it l.h.s.} grows with $n$, one concludes that $\xi \ll 1+h/R_E \approx 4.64$ for large $n$. The allowed region in the parameter space $(n,R)$ is depicted in Fig. \ref{graphpower}.

Analogously to the discussion of the previous section, the range $\ep < 1 \rightarrow R < R_E$ would imply the inclusion of a suitable form factor to model the exchange of spin-2 unparticles within the extended mass distribution of the Earth. The results of Fig. \ref{graphpower} are obtained with no form factor, thus presenting an unrealistic upper bound for the $(n,\xi)$ parameter space: the inclusion of a suppressing form factor would only increase the allowed region for positive $n$ and $\xi < 1$, which is mostly unrestricted.

\subsection{Propagation time delay}

As before, one also computes the additional propagational time delay,

\beq \label{detp} \De t_P = {GM_E \over nc^3} \xi^n \left[ \left({R_E \over R_E + h}\right)^n -1 \right] \approx 1.48 \times 10^{-11} {\xi^n \over n} \left( 0.22^n- 1 \right) ~s ~~. \eeq

\noindent This expression may be simplified by considering the large positive or negative $n$ regimes.

\subsubsection{Large positive $n$}

For a sufficiently large positive exponent $n$, \eq{detp} reads

\beq \De t_P \simeq -{GM_E \over nc^3} \xi^n \approx -1.48 \times 10^{-11} {\xi^n \over n} ~s ~~, \eeq

\noindent and comparison with the time resolution $\De t \sim 10^{-12}~s$ yields the upper bound

\beq {\xi^n  \over n} < 6.76 \times 10^{-2} ~~. \eeq

\noindent If $\xi > 1$, the {\it r.h.s.} grows and this inequality is quickly violated, similarly to the previous paragraph. Hence, one obtains the upper bound $\xi < 1$ for $n \gg 0$, as depicted in Fig. (\ref{graphpower}).

\subsubsection{Large negative $n$}

If $n \ll 0$, \eq{detp} simplifies to

\beq \De t_P \simeq {GM_E \over nc^3} \xi^n \left({R_E \over R_E + h}\right)^n  \approx 1.48 \times 10^{-11} {(0.22\xi)^n \over n} ~s ~~. \eeq

\noindent Equating this to the considered time accuracy $10^{-12}~s$ of Galileo, one obtains the inequality 

\beq {|0.22\xi|^n \over |n|} \ll 6.76 \times 10^{-2} ~~,\eeq

\begin{figure}[]
\centering
\epsfxsize= \columnwidth
\epsffile{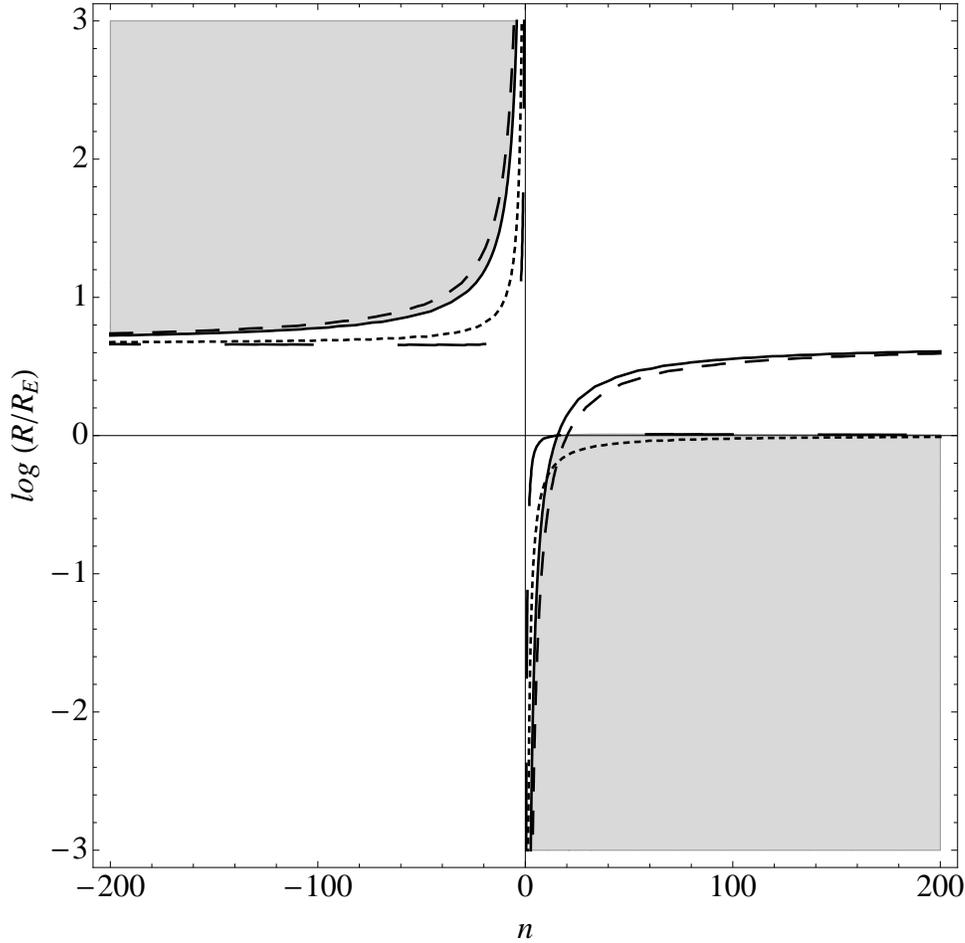}
\caption{Allowed regions (gray) for a Ungravity-type additional force with lengthscale $R$ and exponent $n$, and superimposed limit obtained for a relative frequency precision $\ep_f =10^{-15}$ (full) or $\ep = 10^{-18}$ (short dash), and time resolution $\De t = 10^{-12}~s$ (long dash) or $\De t = 10^{-15}~s$ (dotted).}
\label{graphpower}
\end{figure}

\noindent As before, the {\it l.h.s.} grows with $n$ if $0.22 \xi > 1$; thus, one recovers the previously obtained bound $\xi > (0.22)^{-1} \simeq 4.64$ for large $n$. Notice that this result is independent of the time resolution, since a large negative $n$ regime eventually fulfills this condition, for any given $\De t$. The allowed region for the $\xi$, $n$ parameters is depicted in Fig. \ref{graphpower}, and this asymptotic bound is visible in the negative $n$ axis.

%%%%%%%%%%%%%%%%%%%%%%%%%%%%%%%%%%%%%%%%%%%%%%%%%

%%%%%%%%%%%%%%%%%%%%%%%%%%%%%%%%%%%%%%%%%%%%%%%%%

\section{Conclusions}

In this work, one has assessed the possibility of detecting signals of new physics through the use of the Galileo navigation system. This application could be valuable, as any unexpected new phenomenology could provide further insight into what lies beyond the Standard Model of particle interactions and GR. One has specifically looked at the propagation time delay and frequency shift induced by several corrections to the Newtonian potential: PPN second-order terms, a cosmological constant, a constant anomalous acceleration, a Yukawa addition and a power-law addition.

As it turns out, there is no possibility of using the Galileo satellite navigation system to currently detect PPN second-order additions nor a cosmological constant with the reported value; likewise, a constant acceleration of the order $10^{-9} -10^{-10}~m/s^2$, characteristic of the Pioneer anomaly or MOND, also falls well below detectability.

One finds that the excluded region for the parameter space $(\la,~\al)$ of a Yukawa interaction can be extended with the available precision: the $\la \sim 10^7-10^8$ region is deepened so the allowed range for $\al$ is lowered by about one order of magnitude, to $\al \leq 10^{-8}$.

Moreover, Fig. \ref{yukawa} shows that a possible future generation of satellites, possessing onboard clocks with a relative frequency accuracy $10^{-19}$, might probe a relevant, yet unassessed zone of this parameter space; this kind of clock stability will be eventually available with the onset of quantum time synchronization and subsequent space qualification.

Another interesting result of this study lies in the exclusion of a definite range of values for the characteristic lengthscale of a power-law addition, $R$, namely those lying between $0\leq R \leq 4.64 R_E$. For positive exponents $n$, the region $R < R_E$ is essentially unconstrained (with the allowed region further enlarged by the unaccounted suppression due to the inclusion of a form factor).

Despite of being somewhat incomplete and exploratory, this study serves to show that the Galileo navigation satellite system could prove to be a valuable instrument for improving our understanding of fundamental physics. This result is encouraging, and can be considerably improved in a future generation of Galileo --- particularly if well-thought science objectives are incorporated into the design of the ground segment and satellite constellation.

Furthermore, results concerning the periodic clock rate shift arising due to deviations from a perfectly circular orbit show that a dedicated satellite with a state of the art onboard clock and a highly eccentric orbit could serve to probe yet unconstrained domains of the parameter space of the discussed extensions of GR.

Clearly, this goal is not within the framework set by any global navigation satellite system, which by design aim at a negligible eccentricy orbit. However, one can envisage that a small subset of the constellations could be launched with a sufficiently large amount of fuel (adding to the already provided station-keeping and controlled orbital decay supply), so that its eventual deorbiting could first involve some sort of elliptical path: such dual-use could provide interesting scientific opportunities in a cost-effective way (when compared with a dedicated probe), and serve to highlight the global value of a particular system, beyond its basic navigation purpose.

Although not within the scope of this study, the Galileo satellite navigation system could also prove to be highly valuable in probing the validity of the Local Positioning Invariance (LPI) principle \cite{review}. This postulate, one of the fundamental pillars of General Relativity, asserts that clock rates are independent of their spacetime positions; the available experimental constraints tell us that this invariance holds down to a relative level of $1.4 \times 10^{-6}$ \cite{LPI,LPI2} (see Ref. \refcite{LPI3} for the most recent bounds, and Ref. \refcite{LPI4} for a discussion).

Endowing one or more elements of the Galileo constellation with more precise clocks and providing sufficiently stable communications with ground stations (possibly through a microwave link), an improvement of up to two orders of magnitude on the LPI could be achieved. However, this could be somewhat difficult, given that one does not have direct knowledge of the emitted frequencies, only those of the received signals and the master clocks on the ground.

Indeed, the frequency shift inferred from the comparison between these and the broadcasted timing information could suffice (with increased precision) to ascertain the modifications of gravity here addressed, because one can use the multiple satellites available to correct for some unknown instabilities in the emitters, in a sense obtaining an ``averaged'' signal. However, measuring the LPI would involve comparing the individual space-borne clocks and looking for deviations inferred from the received signals, and it could prove difficult to distinguish actual clock-rate differences due to a violation of the LPI from simple drifts from the original base rate (which should nevertheless be small, for the $10^{-15}$ stability). 

To counteract this issue, an alternative could involve the use of cornercubes on the surface of one or more elements of the Galileo system, thus enabling the use of accurate laser ranging: the obtained redundancy would help to lift this degeneracy and signal ``true'' violations of the LPI. Such an extension of the currently envisioned system could be dubbed {\it Siderius Nuncius}, the Celestial Messenger, after the 1610's historic account of Galileo of the first use of the telescope for astronomical observations. {\it Siderius Nuncius} might constitute a valuable instrument to advance our understanding of the mysteries of the Cosmos.

\section*{Acknowledgments}
This work was developed in the context of the first and second conferences {\it Scientific and Fundamental Aspects of the Galileo Programme} (respectively at Toulouse, 2007 \cite{toulouse} and Padova, 2009 \cite{padova}). The authors thank the organization for the hospitality and Cl\'ovis de Matos for fruitful discussions.

\bibliographystyle{unsrt}
\bibliography{galileo}

\end{document}